\documentclass{svmult}
\usepackage{makeidx}         
\usepackage{graphicx}
\usepackage{bm}
\usepackage{multicol}        
\usepackage[bottom]{footmisc}

\makeindex

\newcommand{\p}{\mathcal{P}}
\begin{document}

\title*{The Enlightened Game of Life}

\author{Claudio Conti}
\institute{
Institute for Complex Systems (ISC-CNR),
Dep. Physics University Sapienza, Piazzale Aldo Moro 2, 00185 - Rome, Italy
\texttt{claudio.conti@roma1.infn.it}
}%
\maketitle

\abstract{
We investigate a special class of cellular automata (CA) evolving in a environment filled by an electromagnetic wave. The rules of the Conway's Game of Life are modified to account for the ability to retrieve life-sustenance from the field energy. Light-induced self-structuring and self-healing abilities and various dynamic phases are displayed by the CA. Photo-driven genetic selection and the nonlinear feedback of the CA on the electromagnetic field are included in the model, and there are evidences of self-organized light-localization processes. The evolution of the electromagnetic field is based on the Finite Difference Time Domain (FDTD) approach. Applications are envisaged in evolutionary biology, artificial life, DNA replication, swarming, optical tweezing and field-driven soft-matter. 
}

\section{Introduction}
The  link between  light and  the development  of  complex behavior  is as  much
subtle  as evident.   Examples  include the  moonlight triggered  mass
spawning of hard corals in the Great Barrier\cite{Levy07}, or the {\it
light-switch  hypothesis} in  evolutionary  biology \cite{ParkerBook},
which   ascribes   the    Cambrian   explosion   \cite{GouldBook}   to  the development  of  vision.  
\\Developing simple mathematical models accounting for the interaction between
a complex system and electromagnetic radiation, 
while stressing self-organization
and collective dynamics, is an interesting and original enterprise.
The basic idea is identifying the most limited set of ingredients including,
on one hand, the electromagnetic origin of light (i.e., not limiting
to ray-tracing and similar techniques) and, on the other hand, a minimal 
description of a complex system affected by illumination.
Such an approach necessarily leads to extremely simplified and un-realistic
theoretical representions, but these are expected to be the starting point
for more complicated treatments for problems like DNA replication
and    accumulation   under    intense    fields
\cite{McCann98,Braun02},    swarming    \cite{CamazineBook},  or 
nonlinear optics of complex soft-materials.
\cite{Yethiraj03,Lumsdon04,Duhr05,Reece07,Conti05,Snoswell06}.
Furthermore practical realizations of these models could be realized
by using light-controlled chemical reactions \cite{costello:026114,PhysRevLett.86.1646}.

Here we  consider the  way the  appearance of
photosensivity affects  the dynamics, the emergent  properties and the
self-organization of a  community of interacting agents, specifically,
of cellular  automata (CA). 
CA are  historically the most
fundamental paradigm  of artificial life  (see, e.g., \cite{Bedau03});
in this work the    renowned   Conway's   Game   of   Life
\cite{Gardner,WolframRMP}  is coupled to  Maxwell's equations. 
This is the  first example  of photosensitive  CA.

\section{The model} 
Our approach is based on two evolutionary problems:
Maxwell equations for the EM field and
the Game of Life (GOL) for the CA.
The latter is represented by an ensemble of squares in 
a 2D box (or {\it cavity}) that can be occupied by a living cell (LC, symbol {\bf 1}), or not (symbol {\bf 0});
each cell has eight neighbors. 
The CA evolution is made by a series of temporal steps obeying the GOL rules:
(i) if a LC has $0,1,4,5,6,7$ or $8$ occupied 
neighbors, it dies (loneliness or overcrowding);
(ii) if a LC has $2$ or $3$ occupied neighbors, it survives
to the next step;
(iii) if an unoccupied cell has $3$ living neighbors, it becomes occupied (self-replication).
In addition we assume the following rule:
(iv) if a LC has collected enough energy from the EM field it survives.
This is modeled by determining the EM energy $\mathcal{E}$ (see below) in the automaton square
and calculating a quantity $\mathcal{P}$, 
which is the fraction of $\mathcal{E}$ that the CA is able to use for life-sustenance. $\mathcal{P}$ obeys the equation
(one for each LC) 
\begin{equation}
\label{e1}
\frac{d \p}{dt}=-\frac{\p (t)}{T}+\frac{\eta}{T} \mathcal{E}(t),
\end{equation}
where $\eta$ is the efficiency ($\mathcal{P}=\eta\mathcal{E}$ in the steady state) and $T$ 
is the dissipation rate, or memory time. 
Indeed we include a power consumption mechanism for the stored EM energy.
We assume that (a) if a LC dies, it looses all its energy,
(b) if $\mathcal{P}$ is greater than a threshold value $\mathcal{P}_{th}$, the LC survives
independently on the number of living neighbors.
\\For a fixed efficiency $\eta$, the CA evolution depends on the available EM energy;
however simple scaling arguments (as outlined below) show that one can use a single
dimensionless parameter the {\it irradiance} $J$. 
If $J=0$ the CA is ``blind'', as the standard GOL, 
conversely, as $J$ increases the effect of the EM field grows.
\subsection{Electromagnetic field equations} 
For a 2D cavity (with perfect mirrors as boundaries), with edge $L$, 
Maxwell equations are written in the TE polarization (i.e. only the fields $E_x,H_y,H_z$ are not vanishing) as
\begin{equation}
\begin{array}{l}
\partial_z H_y -\partial_y H_z=\epsilon_r \epsilon_0 \partial_t E_x\hspace{0.5cm} \partial_{y,z} E_x = -\mu_0 \partial_t H_{y,z}.
\end{array}
\label{e1Max}
\end{equation}
To each element of the CA is associated
a square of material, whose electromagnetic response is
determined by the relative dielectric permittivity
$\epsilon_r$.
Since we are interested in the light-driven CA complex dynamics, we initially neglect any feedback mechanism of the CA
on the field. This is the case of the ``transparent CA'', which corresponds
to take $\epsilon_r\cong1$ and neglect their light absorption. 
We will account for
the nonlinear feedback of the CA on the field in a later section of this chapter.
\\Each CA element is mapped to a square with edge $L_{CA}<<L$.
The energy $\mathcal{E}$ in eq. (\ref{e1})  is given by
\begin{equation}
\mathcal{E}=\int_{cell} \sigma E_x^2 dxdy
\end{equation}
while being $\sigma E_x$ the Ohmic current and $\sigma$
is the conductivity.
For the EM evolution we adopt the Finite Difference Time Domain approach\cite{TafloveBook} and take a monochromatic field with angular frequency $\omega$; 
this is generated by an oscillating dipole placed in the middle
of the cavity, which is switched on for a limited time-slot ($10$ optical cycles).
The corresponding seeding current is sinusoidal with period $2\pi/\omega$ and amplitude $J$.
\subsection{Parameters}
Straightforward rescaling of the relevant equations (\ref{e1}) and (\ref{e1Max})  shows that 
the dimensionless parameters ruling the dynamics are:
$\omega T$, the time constant (i.e. the memory) 
of each CA element expressed in units of the inverse angular frequency;
$\omega t_{CA}$, the time interval between each CA evolutive stage;
$k L_{CA}$, the spatial extension of each CA element in units of the inverse wavenumber $k^{-1}$, with $k=\omega c$ and $c$ the vacuum light velocity;
$k L$, the spatial extension of the cavity;
$\kappa=\mathcal{P}_{th}/\mathcal{E}_\lambda$, the ratio between the threshold energy
for the CA and a reference energy $\mathcal{E}_\lambda=\eta J^2/ \sigma k^2$.
Without loss of generality, 
we can fix $\sigma$ and $\mathcal{P}_{th}$ to 
any value and change $J$ (expressed in dimensionless units hereafter) to modulate the effect of the EM field on the CA dynamics.
Here we choose units such that $\omega=1$, $\omega t_{CA}=1$, $k L_{CA}=1$, $k L=100$ and use $T$ and $J$ as control parameters.
\section{Field and CA evolution}
\begin{figure}
\includegraphics[width=8cm]{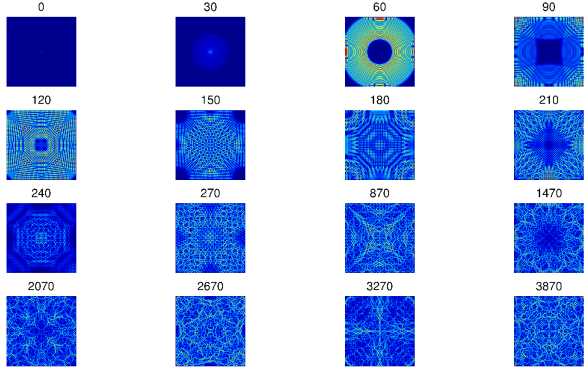}
\caption{(Color online) Snapshots of the field evolution in the cavity for various $t$.
\label{figureem1}}
\end{figure}

\begin{figure}
\includegraphics[width=8cm]{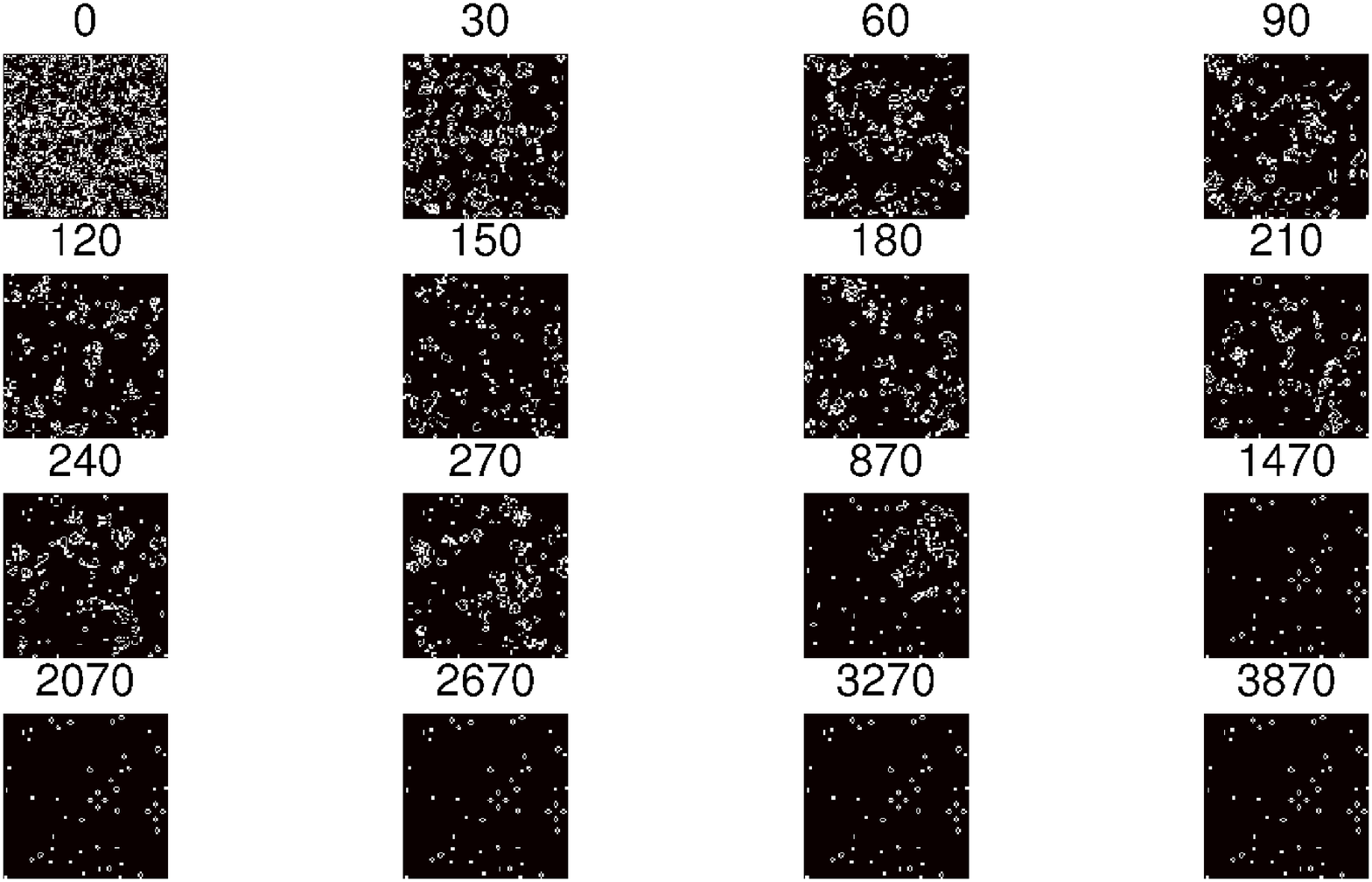}
\caption{(Color online) Snapshots of the CA evolution in the cavity for various $t$ at $J=1$  ($T=10$).
\label{figuresnap1}}
\end{figure}

\begin{figure}
\includegraphics[width=8cm]{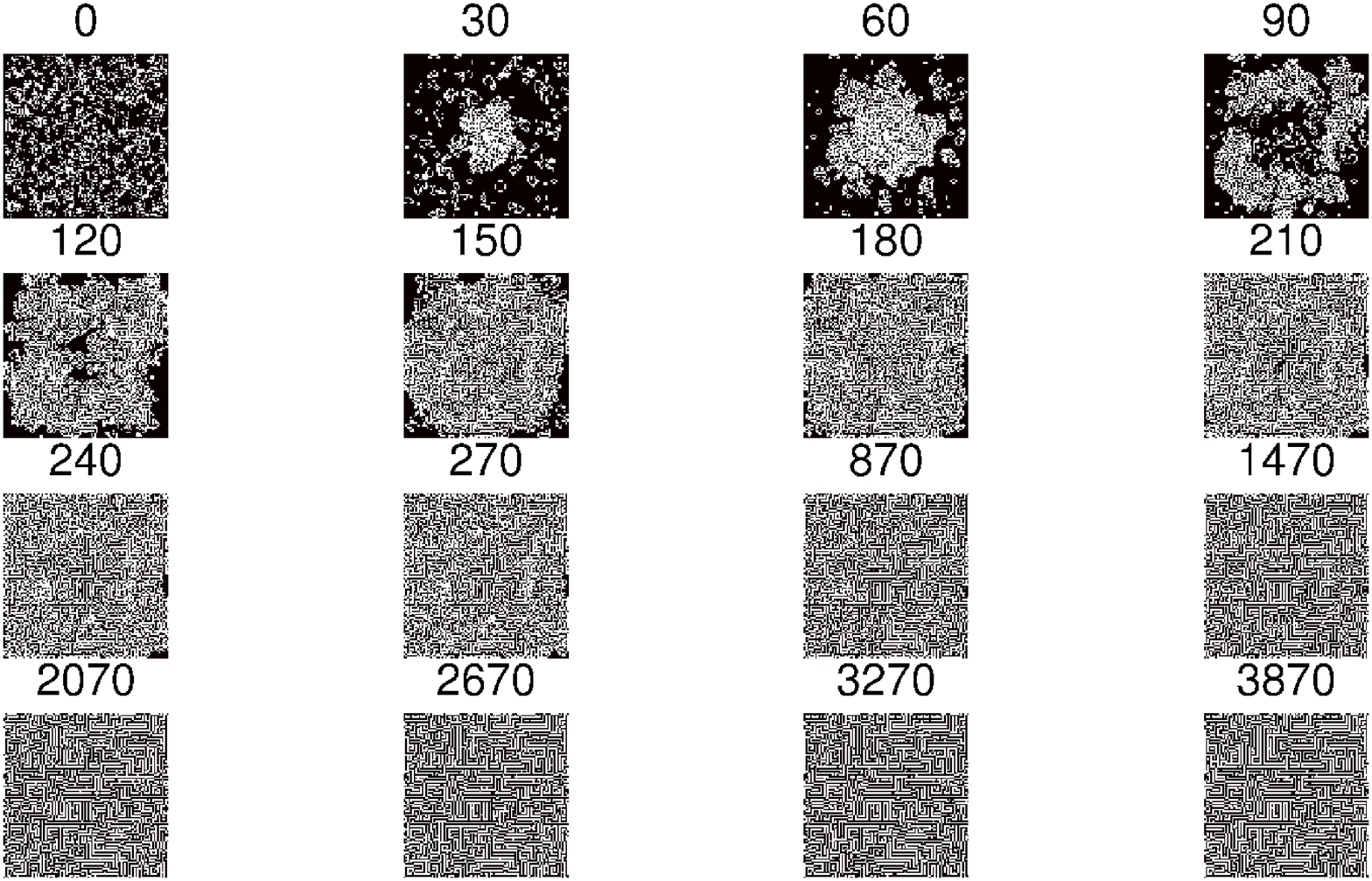}
\caption{(Color online) Snapshots of the CA evolution in the cavity for various $t$ at $J=5$ ($T=10$).
\label{figuresnap5}}
\end{figure}

\begin{figure}
\includegraphics[width=8cm]{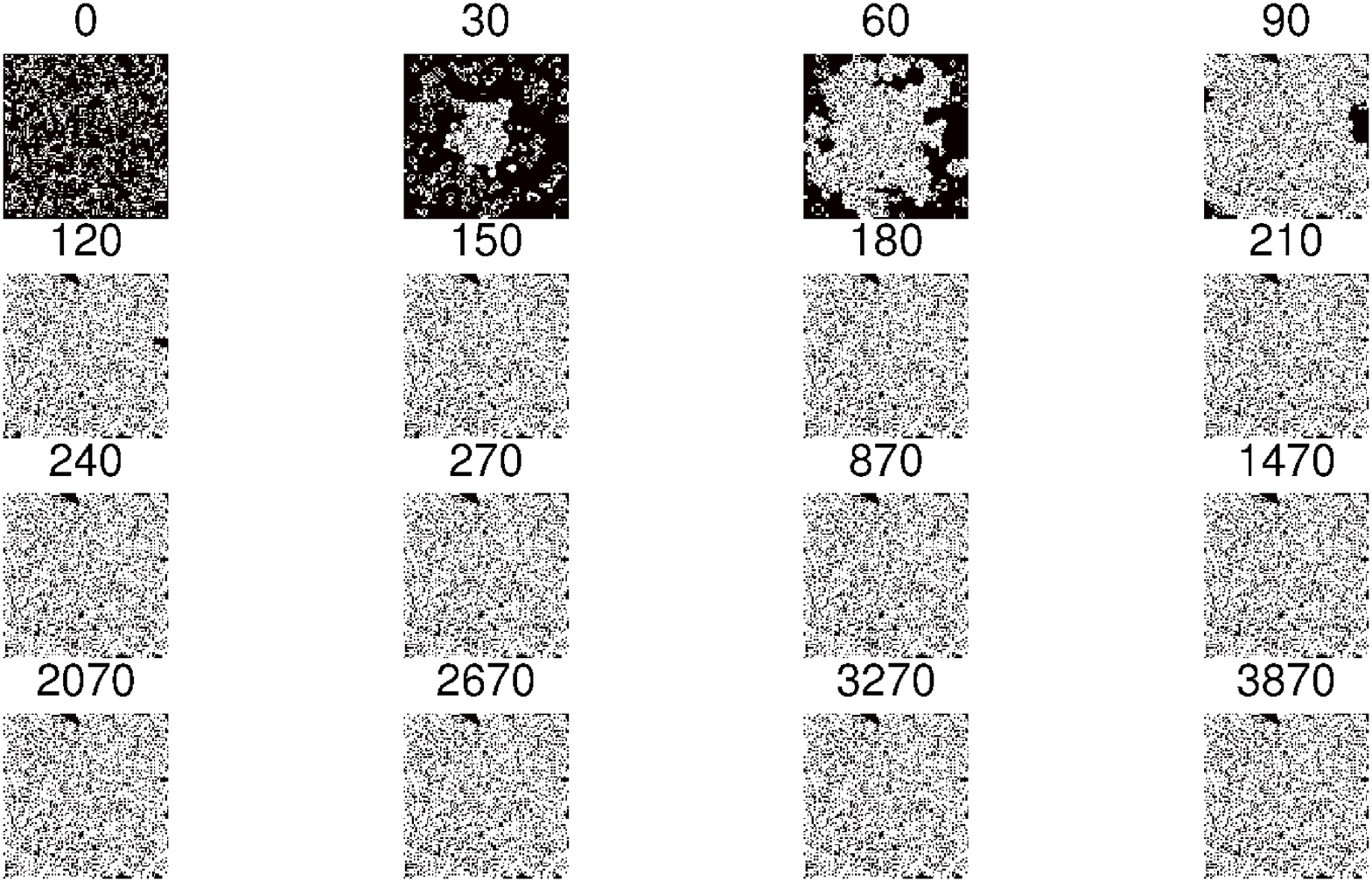}
\caption{(Color online) Snapshots of the CA evolution in the cavity for various $t$ at $J=50$  ($T=10$).
\label{figuresnap50}}
\end{figure}

\begin{figure}
\includegraphics[width=8.3cm]{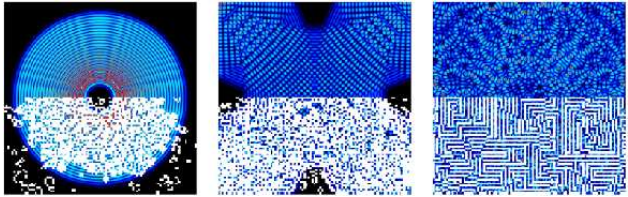}
\caption{(Color online) Photosensitive cellular automata.
Electromagnetic field distribution in the square cavity
at different time instants (from left to right, $t=550,1210,33000$). The contemporary CA distribution
(a small white box for each LC) is superimposed to the field, only its bottom-half is shown. 
(parameters $T=0.1$, $J=1100$)
\label{figureem}}
\end{figure}

We consider the simultaneous EM-CA evolution by starting
from a random configuration of $100\times100$ CA elements in the box.
We show in Fig.\ref{figureem1} various snapshots at different $t$ of the EM field
that, being initially generated in the middle of the structure, progressively fills the cavity.
We show (for $T=10$) in Fig.\ref{figuresnap1} various snapshots of the CA with $J=1$, 
in Fig. \ref{figuresnap5} for $J=5$, and in Fig. \ref{figuresnap50} for $J=50$.
We show in Fig.\ref{figureem} three snapshots of the EM field
in the cavity with the corresponding CA distribution. 
In the early stages the CA is disordered, while a complex pattern appears at long times;
this is largely affected by the degree of photo-sensitivity determined by the parameter $J$.

\begin{figure}
\includegraphics[width=8.3cm]{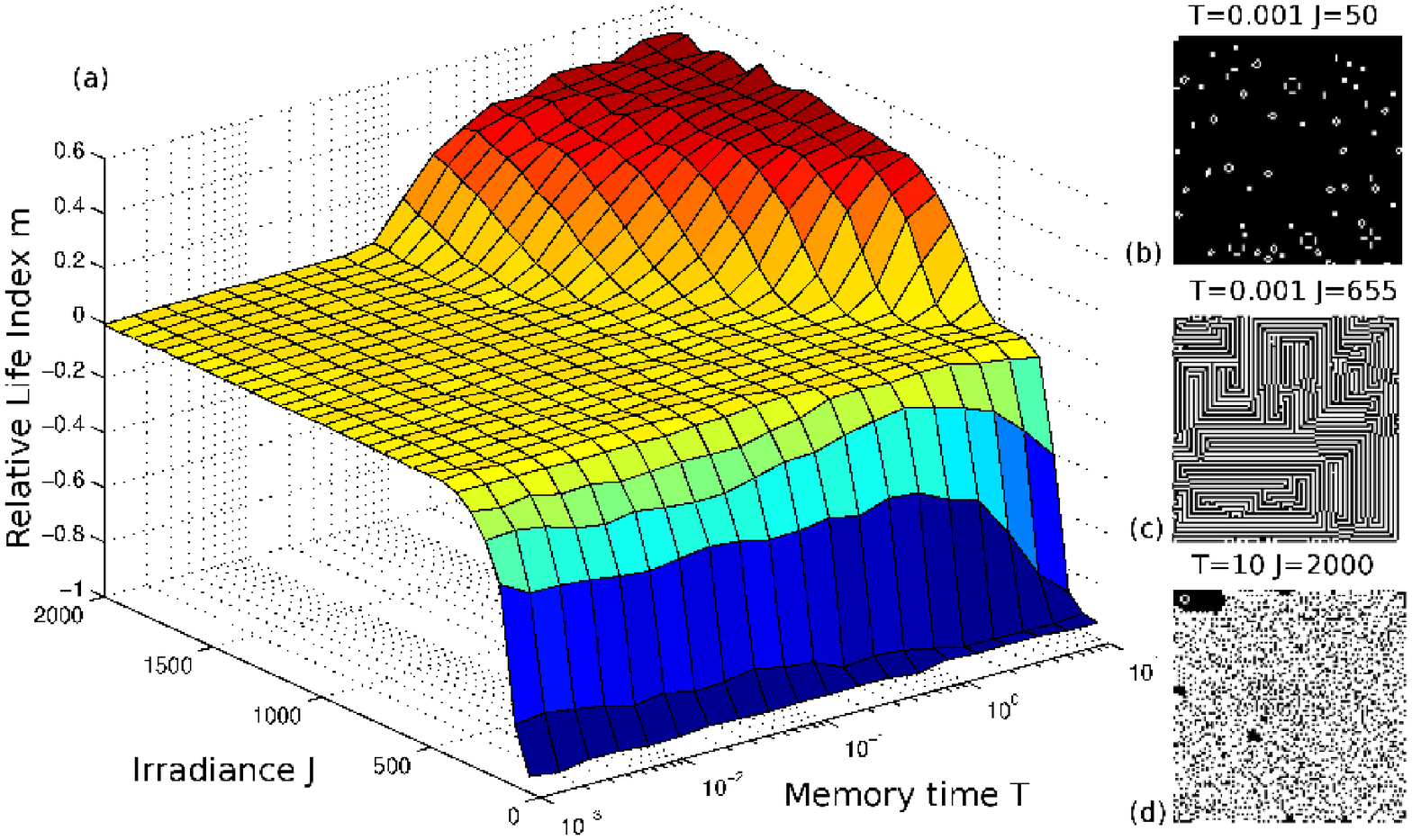}
\caption{(Color online) Relative Life Index Vs irradiance and memory.
The panels on the right show three long-time ($t=4000$) CA configurations with
the corresponding parameters.
\label{number1}}
\end{figure}

\section{Stationary properties of the CA}
The various CA phases can be characterized by the number of LC; 
this is quantified by the relative life index (RLI) $m$, which is calculated
by assigning an Ising spin $\sigma$ with value ``-1'' to {\bf 0} and a ``+1'' to {\bf 1}.
\begin{figure}
\includegraphics[width=8.3cm]{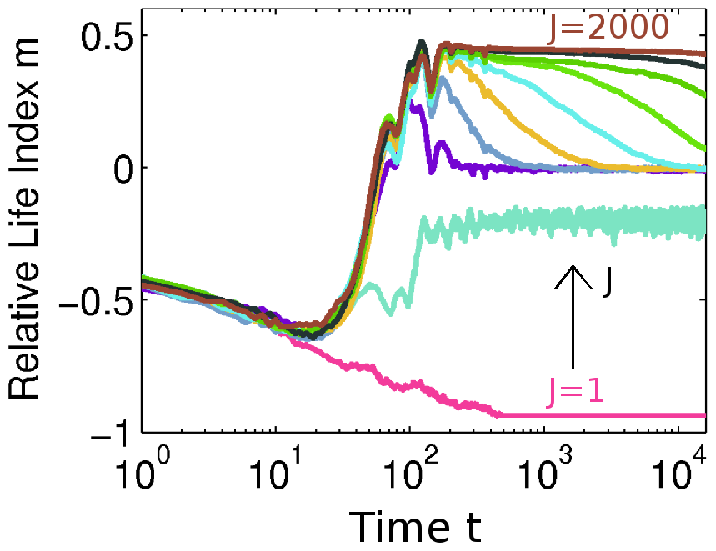}
\caption{(Color online) Life index Vs time for increasing irradiance (here $T=1$).
\label{figure3}}
\end{figure}
The RLI is the average value of $\sigma$ over all the CA.
A configuration with many {\bf 0}s exhibits negative values of $m$
($m=-1$ for all {\bf 0}), while $m= 1$ for all LC.
In the 3D plot of $m$ Vs $T$ and $J$, three regions
can be identified (Fig.\ref{number1}a). At very low irradiance 
(small efficiency or low EM intensity) the final population is organized as in the standard 
GOL (Fig.\ref{number1}b):
it is characterized by small-size unconnected communities of LC ({\it blind} phase).
Two additional regimes are found while increasing $J$:
(i) a {\it glassy} phase (where $m=0$) with regular domains
separated by various defects (Fig.\ref{number1}c); 
(ii) a region where the CA is frozen 
in a large disordered configuration with $m>0$ (Fig.\ref{number1}d).
In the glassy phase (plateau in Fig.\ref{number1}a),
the CA is not sensible to any increase of $J$.
In this regime the EM field sustains a large amount of LC, but their number
is frustrated by the internal self-organization.
This is true as far as the region with $m>0$ is entered, where an explosive growth of the LC with the
irradiance (and the memory time) is found; this is the {\it evolved phase}. 
The existence of this transition is a result of the competition between 
the GOL rules and the effective employment of the EM energy for life-sustenance. 
\section{Dynamics}
In figure \ref{figure3} we show the time evolution of $m$ for increasing $J$ at a fixed memory. 
Starting from the same random CA, different histories are determined
by EM field. In the blind phase, the CA rapidly evolve 
to a small number of LC separated in isolated communities (see \cite{WolframRMP}).
At sufficiently high fluence, the RLI overshoots
and then decays to zero. This implies that the EM field
favors the life of a large number of CA;
however this is sustainable for long times only at very high
irradiance (evolved phase); in the other regimes the population steadily decays to $m=0$.
\begin{figure}
\includegraphics[width=8.3cm]{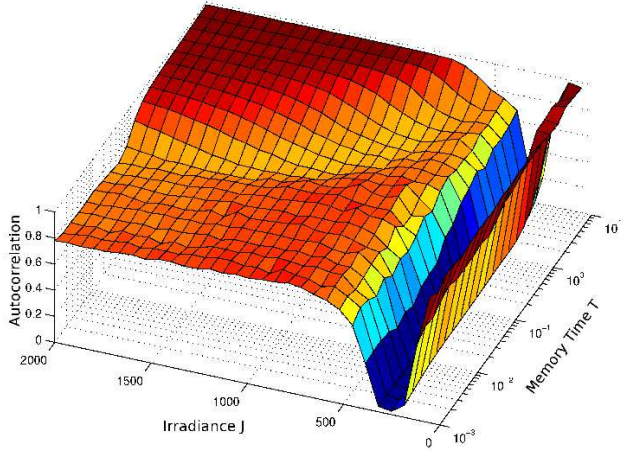}
\caption{(Color online) Autocorrelation at $\tau=4000$.
\label{auto1}}
\end{figure}
\\
\noindent Figure \ref{auto1} shows the autocorrelation function averaged over all the CA:
\begin{equation}
\phi(\tau)=\langle \int  \sigma_i(t+\tau)\sigma_i(t) dt \rangle_{CA},
\end{equation}
which is normalized such that $\phi(0)=1$.
When increasing the strength of the interaction with the EM field from the blind GOL ($J\cong0$), 
the CA first display a disordered dynamical phase ($m\cong-1$ and $\phi\cong0$), then a glassy region (where $m=0$ and $\phi\cong0.8$). 
In the evolved phase ($m\cong1$), $\phi\cong1$ denotes a frozen configuration.
\section{Self-healing after a catastrophic event}
\begin{figure}
\includegraphics[width=8.3cm]{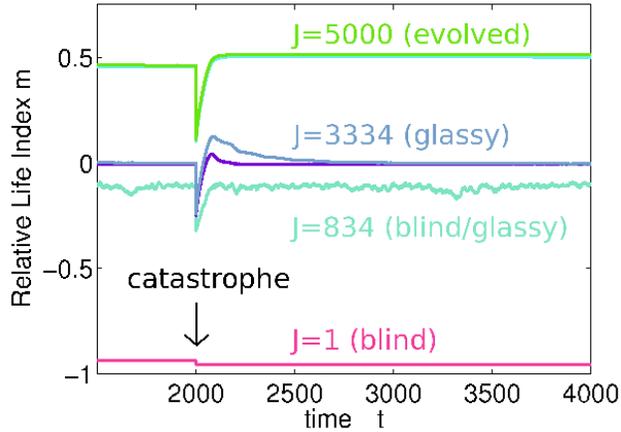}
\caption{(Color online) Self-healing after a catastrophic event.
Relative life index Vs time for various irradiances $J$ 
(not all values of $J$ for the reported lines are shown) in the presence
of the abrupt killing of the living cells in the middle area of the box
(at $t=2000$, here $T=1$).
\label{catastrophe}}
\end{figure}

We then consider the reaction of the photosensitive CA to ``catastrophic events''.
We let the system evolve to a stationary state
and then we ``kill'' all the cells occupying a square in the middle of the box
and with size $L/2$ (see figure \ref{catastrophe}).
In the blind phase the system does not react to this event, and the RLI is reduced.
Conversely, in the 
glassy and in the evolved phases, the system rapidly restores
the number of LC; 
at high fluences this is also accompanied by an overshoot
of the RLI, which decays to zero in the glassy phase.
\begin{figure}[!t]
\includegraphics[width=8.3cm]{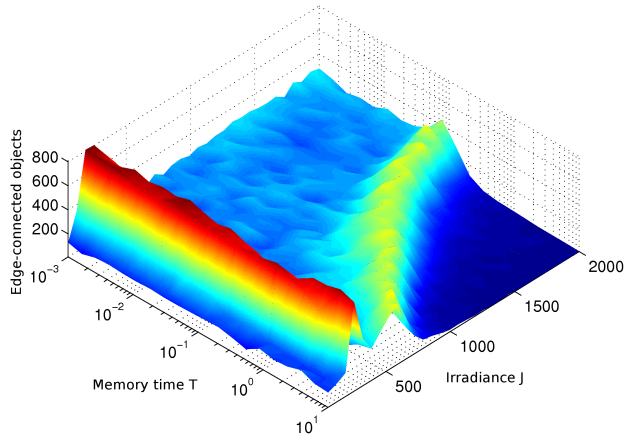}
\caption{(Color online) Number of edge-connected regions Vs memory time and irradiance.
\label{conn4}}
\end{figure}
\begin{figure}
\includegraphics[width=8.3cm]{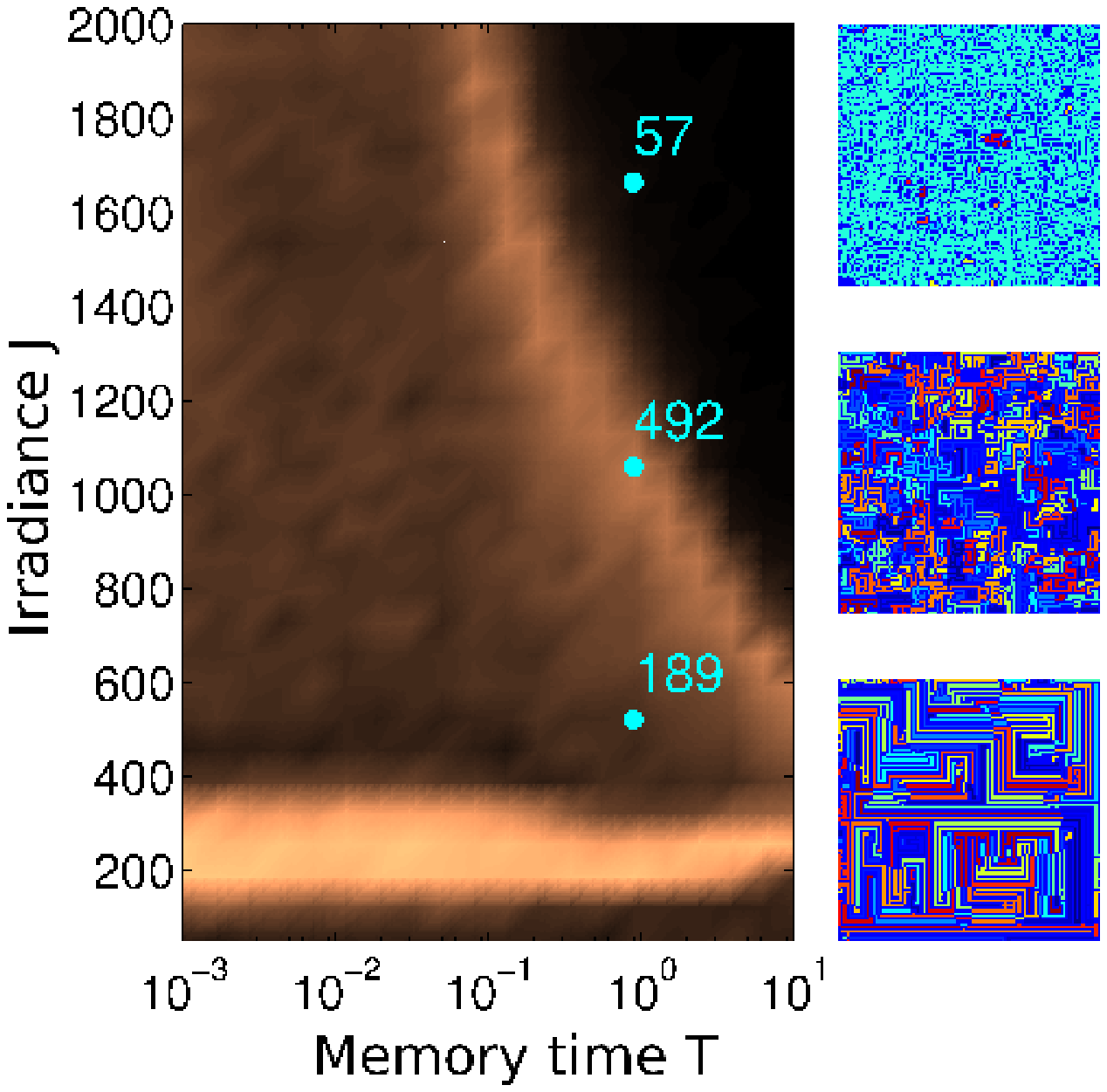}
\caption{(Color online) Colormap of the number of edge-connected regions;
the three dots correspond to the CA panels on the right;
connected regions are discriminated by different colors.
\label{figure7}}
\end{figure}
\section{Topology and self-organization}
To characterize self-organization we count the number of edge-connected objects (or {\it communities})
in the large time ($t=4000$) CA configuration .
Figure \ref{conn4} shows a three-dimensional plot of the number of edge-connected regions versus the irradiance and
the memory time.
In the blind phase, one has a large number of unconnected very-small communities with $m\cong-1$. 
In the glassy phase, many connected regions with $m=0$ are found.
In the evolved phase ($m>0$), the CA is organized into a small number of large communities.
Specific transition regions can be identified (peaks in Fig.\ref{conn4}, brighter lines in Fig. \ref{figure7})
and these are characterized by tiny ranges of the parameters with
a huge number of small unconnected communities.
Indeed, the transition from the glassy phase to the evolved one is driven
by the breaking of the almost regular domains (see panels in Fig.\ref{figure7}).
Correspondingly the number of communities first increases
(in Fig.\ref{figure7} they change from $189$ to $492$) and then rapidly decreases 
in correspondence to the formation a large amorphous but connected CA at high irradiance.
\section{Introducing a genetic code and inheritance}
\begin{figure}
\includegraphics[width=8.3cm]{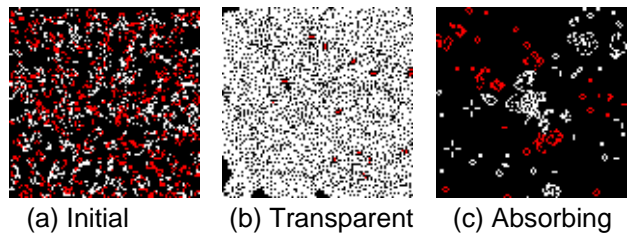}
\caption{(Color online) 
Photosensitive CA with genetic code.
(a) Initial configuration with 
same number of blind (red, darker) and photosensitive (white) LC;
(b) final configuration for the transparent CA;
(c) as in (b) for the absorbing CA
($T=0.1$, $J=50$, $t=1000$).
\label{figuregenetic1}}
\end{figure}

One can argue if the photoreception ability can favor some
evolution of the CA toward novel species.
The simplest mechanism to be considered is that based 
on natural selection, such that one assume that 
a ``gene'' responsible for photoreception is randomly
distributed among the LC. Those LC not displaying
such a gene are blind (they obey to the simple GOL rules);
the others behave as described above and feel the presence of the EM field.
When a new LC is born from the three neighbors [GOL rule (iii) above],
it inherits the photoreceptive gene if this is present
in two or three of the parents, otherwise it is blind.
We find that, 
as far as no EM field is present, the photosensitive population balances
the blind one. When the EM field is introduced, the PLC rapidly supersede the BLC. Figure \ref{figuregenetic1}b shows 
the large-time state of the CA when starting from a balanced configuration
(Fig. \ref{figuregenetic1}a).
Letting $n_p$ the number of PLC, 
and $n_b$ that of the BLC, we show
the ratio $g=(n_p-n_b)/(n_p+n_b)$ in Fig.\ref{figuregenetic2}
(left axis) Vs time. In the transparent case, $g$ rapidly 
reaches the unity.
\section{Energy dissipating CA}
\begin{figure}
\includegraphics[width=8.3cm]{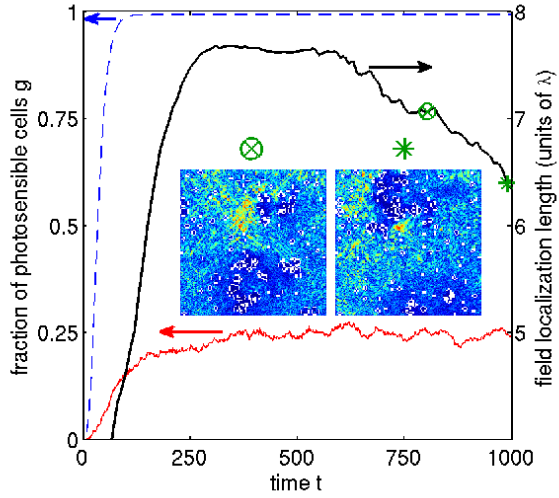}
\caption{(Color online) 
Left scale: fraction of PLC Vs time averaged over $100$
initial balanced random condition for 
transparent (dashed) and absorbing (full line) CA.
Right scale: field localization length Vs time 
($100$ initial random configuration)
for the absorbing CA. The insets show the EM field distributions
(for the absorbing case with the CA superimposed) at two instants 
($\otimes$ and $\ast$) unveiling the self-organized dynamic 
wave localization ($J=50$, $T=1$).
\label{figuregenetic2}}
\end{figure}

The ``gene'' selection process implies that PLC are favored in the presence of an
external EM field. The situation however is different if one 
takes into account the fact that the PLC absorb energy:
as their number grows the life-sustaining field
is reduced and the selection process is frustrated.
In figure \ref{figuregenetic1} 
we compare the transparent case with the absorbing one 
($\sigma=10^5$\,S\,m$^{-1}$ for the CA material).
In the presence of dissipation 
the fraction of photosensitive agents is reduced,
however, surprisingly enough, it stays constant with time after an initial build-up transient.
\\ When considering the snapshots of the EM profile during
the CA evolution, one readily realizes that, at variance with
transparent case (where the EM wave is de-localized in the entire
cavity, see Fig. \ref{figureem}), the field displays a certain degree of
localization. Indeed regions with high intensities appears, 
circumvented by various LC (insets in Fig.\ref{figuregenetic2}). 
The effect can be quantified by calculating the EM localization
length $l_{EM}$ (see, e.g., \cite{Gentilini09}),
 reported in Fig.\ref{figuregenetic2} (right axis).
Notably, after a transient over which the field fills the cavity
(up to $t\cong250$), $l_{EM}$ 
starts to decrease with time, while $g$ stays constant.
As the PLC dissipate energy,
the CA is able to localize light 
(insets in Fig. \ref{figuregenetic2}) in order to preserve the intensity level. 
\section{Conclusion}
Within the proposed model of photosensitive (artificial) life,
one finds that the development of photoreception largely affects
not only the number of living automata but also their organization.
If the storage time is too small, the population cannot grow;
it wastes energy more quickly than the time needed to collect it.
Conversely, an explosive growth is found at the expense of large-scale self-organization, which appears
only after a critical degree of photosensibility has been developed.
Self-healing abilities after catastrophic events and dynamical hierarchies are triggered
by the EM radiation.

When introducing a genetic-like competition between photosensitive and blind CA,
the former are favored by the irradiation. 
If the CA energy dissipation is included,
the highly nonlinear EM-CA system
results into a self-organized field localization effect, such that the EM localization
length decreases with time in order to keep constant the number of photosensitive agents.

The proposed model shows that the competition between
the internal rules of a complex system and the development of new abilities (as vision)
nurtures abrupt evolutive steps and collective behavior.

\section{Acknowledgments}
We acknowledge support from the
INFM-CINECA initiative for parallel computing. The
research leading to these results has received funding
from the European Research Council under the European
Community's Seventh Framework Program (FP7/2007-
2013)/ERC grant agreement n.201766.

\printindex
\end{document}